# String Theory, Universal Mind, and the Paranormal *

Brian D. Josephson
Department of Physics, University of Cambridge

Cavendish Laboratory, Madingley Rd, Cambridge CB3 0HE, U.K.
http://www.tcm.phy.cam.ac.uk/~bdj10

**ABSTRACT**
A model consistent with string theory is proposed for so-called paranormal phenomena such as extra-sensory perception (ESP). Our mathematical skills are assumed to derive from a special ‘mental vacuum state’, whose origin is explained on the basis of anthropic and biological arguments, taking into account the need for the informational processes associated with such a state to be of a life-supporting character. ESP is then explained in terms of shared ‘thought bubbles’ generated by the participants out of the mental vacuum state. The paper concludes with a critique of arguments sometimes made claiming to ‘rule out’ the possible existence of paranormal phenomena.

## 1. Introduction

Critics of claims of the paranormal, e.g. Deutsch (2001), have declared extrasensory perception (ESP) or other paranormal phenomena to be ‘nonsense’. Such absolutist positions give little weight to the experimental evidence (Radin 1997) in support of the reality of such processes, and seem naive given the range of imaginative proposals concerning the nature of reality currently being put forward for serious consideration by conventional physicists. One important advance has been the superseding of the so-called Standard Model as a fundamental theory of nature by string theory (http://superstringtheory.com), where the Standard Model features merely as a subset of the set of permitted possibilities. As Carr (2001, 2003) (whose approach is centred on the alternative Randall-Sundrum picture) has suggested, such a change in perspective opens up new possibilities in science, including the possibility of accommodating paranormal phenomena within physics. In the following a number of concepts are combined, each in essence consistent with accepted ideas, resulting in a qualitative explanation for ESP, with the promise of an eventual clear cut basis for understanding paranormal phenomena in general.

## 2. A separate mental reality

A key assumption we make is one which, while it has no clear connections with experimental physics, does make contact with a position that was advocated by mathematicians such as Gödel (Davis and Hersh 1981, Penrose 1994). This is the idea that some aspects of mentality involve a realm of reality largely, but not completely, disconnected from the phenomena manifested in conventional physics. The idea of a disconnected realm does have precedents, for example in the way two of the fundamental forces (the strong and weak forces) play no role in large areas of physics and chemistry, whilst in other contexts they have a very important part to play. Next note that string theory, involving as it does spaces having more dimensions than the usual three, and also a non-unique vacuum state (and according to Susskind (2003a, b), a very large number of such states), is consistent with there being such a 'separate realm', in a way that the Standard Model, with its unique vacuum state contained within a limited number of spatial dimensions, did not.

The point in regard to mathematical thinking, which motivates our model, is the following. Consider first of all what the brain does in visual perception. Here the primary information from the visual receptors goes through various levels of processing until it ends up as a high-level representation of the content of the visual field. It is not unreasonable to identify mathematics as a similar process, except that higher levels of abstraction are involved in this case. With the visual case, the mechanics are straightforward: the visual field typically contains for example edges, for which abstraction a dedicated neural system has evolved, related to our ability to perceive edges. It is hard to see why we should have such ready access to higher mathematical abstractions having little connection with experience (Penrose 1994). One resolution of the problem would be for mathematical concepts to be in some way 'in the physics', rather than being emergent properties of brains. In case it is felt that such a drastic solution is not necessary to explain our ready access to mathematical ideas, and that neural networks can provide an adequate explanation, a stronger argument for the existence of some kind of Platonic realm can be made on the basis of the aesthetic aspect of music (Josephson and Carpenter 1996).

So far, in shifting the locus of mathematical thinking (and music?) to another realm, we have only replaced one mystery by another. But why should such a realm exist at all? The explanation we provide is of a biological character, taking account of the fact that information processing is an essential component of biological functioning, but with only very specific informational processes having a life-supportive character. While it is commonly taken that the informational processes involved are mediated by ordinary physical means, it is not a logical necessity that this should be the case. Some informational processes in an organism are specialised to the nature and circumstances of the organism concerned, but some have a more abstract and universal character, and so could be mediated by a quite different system with which individual organisms would interact.

Next we observe that a form of proto-life, defined as fluctuation patterns surviving longer than typical patterns do, can be hypothesised as occurring at the Planck scale, evolution of such life being expected to involve evolution of the accompanying informational systems also. We get to the proposed model by supposing that the ordinary physical component and the informational component can evolve separately. and that the informational component can even survive the creation and destruction of individual universes, remaining as an ever-present background with which new universes, Planck scale fluctuations and more developed life forms can all beneficially interact. Assuming an indefinitely extended time scale, the most persistent part of the informational background can evolve indefinitely, so that its dynamics might come to include features corresponding to mathematical concepts and operations as well.

This idea can be fruitfully connected with anthropic ideas, particularly taking the point of view of Susskind (2003a, b), who explains the way our universe seems to be mysteriously fine tuned to develop in such a way that life is possible in terms of it being only one of a vast number of coexisting universes, a small proportion of which have such a property, one of which we find ourselves occupying. Our speculations can be seen as the application of a similar idea to the informational aspect of life.

While Susskind treats life as a passive occupant of whatever universe can permit it to develop, our proposals see life in a more general light, able to shape its environment in a partnership with it, in a manner analogous to the proposals of Lovelock (1995) (the Gaia hypothesis, for which there is now considerable supportive evidence), to the effect that life may be able to interact cooperatively with its environment, discovering how to operate upon it to its best advantage.

**3. A model for ESP**

We need to add another piece of detail to our model. In order that it can model individual thought, we suppose that individual life forms can perturb the background state so as to create a localised 'thought bubble', tied to the individual concerned. This suggests that the vacuum state involved is close to a phase transition, so that an appropriate perturbation can create a domain with a different kind of order to that of the vacuum.

Assuming the validity of the scenario that has been described, the picture proposed can be adapted to account for the phenomena we set out to explain, namely telepathy or ESP. In the first, the grounds for the existence of such a process can be taken to be the advantages that might be conferred in certain situations if two life forms could in some way share their mental states (there could also be accompanying disadvantages, the significance of which will become clear later). It is natural to postulate, in this case, that a *shared* 'mental bubble', whose contents are available to both life-forms, is involved. We assume, as would need to be assumed generally in the model, that the state of this bubble plays the role of information that is meaningful in the context and, by virtue of this, usable by the connected systems.

The physics involved in the ‘sharing’ that has to be assumed in the above can be clarified by means of an analogy based on the Mössbauer effect, which is a phenomenon involving the decay of radioactive nuclei embedded in a crystal (Mössbauer 1961). In a certain fraction of cases, depending on parameters such as the decay energy and the temperature, the recoil from such a decaying nucleus is in effect transmitted to the crystal as a whole rather than generating activity in the vicinity of the decay. These ‘no local recoil’ processes involve a certain subset of all possible final states of the system, for which, as quantum mechanics allows, the state of the lattice vibrational system (phonons) is unchanged by the decay. This somewhat esoteric possibility suggests a mechanism, dependent on analogous constraints upon the possible states of the thought bubble, that could fit our requirement of a system state being shared by two individuals as in the ESP situation.

## 4. Countering the critics

The problem any such analysis has to face is that of explaining how it is that, if such a mechanism for ESP or other paranormal processes exists, these processes manifest themselves only in very specific ways, and in ways that are not readily controllable. This should not be seen as an insuperable objection, since other phenomena (e.g. those involving the weather), have similar features. The point to bear in mind is that in the biological realm the phenomena that manifest are governed not only by what is physically possible, but also by which of those physically permitted possibilities are likely to be of overall benefit to the organism concerned. In the ESP context, an undifferentiated sensitivity to the thoughts of all other people, as would result from the uncontrolled sharing of thought bubbles, would tend to be disadvantageous rather than of benefit, leading to the individual being overwhelmed by thoughts of others. Most of these would be merely distracting, and interfere with constructive activity. The right way to think about ESP is therefore to see it as a slowly developing phenomenon for a given individual, and one which may not develop at all if conditions are unfavourable. We see from this analysis that the frequently made counter-argument to the existence of ESP, that if it were possible it would have such a survival value that we would all evolve to be very good at it, is based on a misleading concept of what would be involved.

A related problem is the one raised by Weinberg (1993), who asks what possible physical signal could move distant objects and yet have no effect on scientific instruments? Such a question ignores the possibility that there might be a threshold for psychokinetic effects. A similar argument would lead one to be equally sceptical of claims that the heat of the sun can induce chemical reactions (i.e. burning) in a piece of paper, analogously something that happens only under special circumstances (e.g. using a magnifying glass to focus the sun’s rays on to a spot on the paper), the amount of burning under normal conditions being negligible.

The moral to draw would seem to be that one should not be too ready to dismiss paranormal phenomena on the basis of apparently plausible arguments; as ’t Hooft (2001) has said in a slightly different context, plausible arguments come with their own ‘small print’, viz. assertions to the effect that assumptions that seem reasonable to their authors may be violated in the real world.

## 5. Concluding comments

This work was motivated primarily by the perception that the arguments commonly made against the possible existence of paranormal phenomena are not well-founded, suggesting a need to discover how they might be accommodated within conventional science. Proposals with this aim have been made previously, based upon Bohm’s causal interpretation of quantum mechanics (Josephson and Pallikari-Viras 1991, Valentini 1991), but the fact that the causal interpretation of quantum mechanics has not developed in ways relevant to current scientific concerns suggests it may be more fruitful to look elsewhere for ideas. The present paper is the outcome of such an investigation. Clearly, it is at best a sketch of a theory, since the arguments are of a very qualitative character, but this qualitative sketch brings to light a number of specific issues whose resolution may provide the basis for a more complete account of the phenomena.

## 6. Supplementary remarks (added after submission of paper for Proceedings)

Susskind's arguments suggest that reality may be much more complex than has normally been assumed. Further changes in fundamental science (which may include consideration of the influence of life) may be required to address this complexity. Since our proposals (such as thought bubbles emerging from some kind of background) do not involve the precise details of string theory, they may survive any such changes that fundamental science may undergo.

## 7. Acknowledgements

The author is indebted to Dr. Fotini Pallikari for many illuminating discussions concerning the nature and mechanisms of ESP. No funding from counter-innovative sources was involved with the preparation of this paper.

Note on categories for this paper: I believe that this paper would be most appropriately listed in gr-qc or hep-th, but the automatic system transferred the initial posting from hep-th to physics, subsequently blocking attempts to crosspost to either of the above lists. Requests to the archive for this situation to be remedied have, to date, met with no response. Comments people may have on this state of affairs could be directed to the moderators (for details see http://arXiv.org/uploads), or to members of the advisory board (see http://arXiv.org/ad-board.html for a list).